\documentclass[pra,aps,amsmath,amssymb,amsfonts,twocolumn,nofootinbib,longbibliography]{revtex4-1}
\usepackage{amssymb}
\usepackage{bm,mathrsfs}
\usepackage{graphicx}
\usepackage{epsfig}
\usepackage{amsmath,bbm}
\usepackage{amsfonts,amssymb}
\usepackage{times}
\usepackage{verbatim}
\usepackage[sort&compress]{natbib}
\usepackage{amsmath}
\usepackage[colorlinks,breaklinks,linkcolor=blue,anchorcolor=blue,citecolor=blue,urlcolor=blue,dvipdfm]{hyperref}

\begin{document}

\title{One-step implementation of Toffoli gate for neutral atoms based on unconventional Rydberg pumping}

\author{H. D. Yin}
\affiliation{Center for Quantum Sciences and School of Physics, Northeast Normal University, Changchun 130024, China}
\affiliation{Center for Advanced Optoelectronic Functional Materials Research, and Key Laboratory for UV Light-Emitting Materials and Technology
of Ministry of Education, Northeast Normal University, Changchun 130024, China}

\author{X. X. Li}
\affiliation{Center for Quantum Sciences and School of Physics, Northeast Normal University, Changchun 130024,  China}
\affiliation{Center for Advanced Optoelectronic Functional Materials Research, and Key Laboratory for UV Light-Emitting Materials and Technology
of Ministry of Education, Northeast Normal University, Changchun 130024, China}

\author{G. C. Wang}
\email{wanggc887@nenu.edu.cn}
\affiliation{Center for Quantum Sciences and School of Physics, Northeast Normal University, Changchun 130024, China}
\affiliation{Center for Advanced Optoelectronic Functional Materials Research, and Key Laboratory for UV Light-Emitting Materials and Technology
of Ministry of Education, Northeast Normal University, Changchun 130024, China}

\author{X. Q. Shao}
\email{shaoxq644@nenu.edu.cn}
\affiliation{Center for Quantum Sciences and School of Physics, Northeast Normal University, Changchun 130024, China}
\affiliation{Center for Advanced Optoelectronic Functional Materials Research, and Key Laboratory for UV Light-Emitting Materials and Technology
of Ministry of Education, Northeast Normal University, Changchun 130024, China}

\begin{abstract}
Compared with the idea of universal quantum computation, a direct synthesis of a multiqubit logic gate can greatly improve the efficiency of quantum information processing tasks. Here we propose an efficient scheme to implement a three-qubit controlled-not (Toffoli) gate of neutral atoms based on unconventional Rydberg pumping. By adjusting the strengths of Rabi frequencies of driving fields, the Toffoli gate can be achieved within one step, which is also insensitive to the fluctuation of the Rydberg-Rydberg interaction. Considering different atom alignments, we can obtain a high-fidelity Toffoli gate at the same operation time $\sim 7~\mu s$. In addition, our scheme can be further extended to the four-qubit case without altering the operating time.
\end{abstract}

\maketitle
\section{Introduction}
Neutral atoms excited to Rydberg states with large principal quantum numbers have been a potential platform for quantum information processing~\cite{Saffman2010RMP}. The Rydberg states not only have a long lifetime, but also interact with each other through strong long-range Rydberg-Rydberg interactions~(RRI), i.e, dipole-dipole or van der Waals interaction. These strong interactions can prohibit the excitation of adjacent atoms once one atom has been excited, which is called the Rydberg blockade effect~\cite{Jaksch2000PRL,Lukin2001PRL}. Since Jaksch $et~al.$ put forward the scheme of quantum gate built on Rydberg blockade, many technologies for implementing quantum gates based on Rydberg atom have been proposed, such as adiabatic passage~\cite{Moller2008PRL,Beterov2013PRA,Rao2014PRA,WHZ2017PRA,Petrosyan2017PRA,Saffman2020PRA}, electromagnetically induced transparency~\cite{Muller2009PRL,Gorshkov2011PRL,ZYL2011PRA,Paredes2014PRL,Baur2014PRL,Das2016PRA,Lahad2017PRL,Tiarks2019NaturePhysics}, Rydberg dressing~\cite{Keating2013PRA,Petrosyan2014PRL,Muller2014PRA,Keating2015PRA,Lee2017PRA,SXF2018PRA,Mitra2020PRA} and F\"orster  resonance~\cite{Tiarks2014PRL,Ravets2014NaturePhysics,Ravets2015PRA,Beterov2015PRA,Gorniaczyk2016NatCommun,Beterov2016PRA,Beterov2018PRA,HXR2018PRA}.

Another effect called Rydberg antiblockade has been proposed theoretically~\cite{Ates2007PRL} and demonstrated experimentally~\cite{Amthor2010PRL}. In contrast with Rydberg blockade, Rydberg antiblockade can directly excite two Rydberg atoms from the ground states to the Rydberg states without single-excited Rydberg state. This kind of excitation process provides a method for rapid construction of quantum gates, it can be used to construct controlled-phase gate~\cite{SSL2016PRA,SSL2017PRA_1,SSL2017PRA_2,SSL2018PRA,SSL2018CPB,ZXY2019JOSAB,WJL2020PLA,WJL2020OL} and controlled-not gate~\cite{SSL2017PRA_2,WJL2020PLA}. However, it is difficult to exactly control the distance between the two Rydberg atoms in the experiment to meet the condition $U=2\Delta$, where $U$ denotes the van der Waals interaction and $\Delta$ is the single-photon detuning parameter. Therefore, once the parameter relationship is disturbed, the scheme will fail due to the rapid decline of fidelity.
Recently, our group proposed an unconventional Rydberg pumping (URP) mechanism ~\cite{LDX2018PRA}, which is different from Rydberg blockade and Rydberg antiblockade. By driving the same ground state of each atom with a dichromatic classical fields, the evolution of the atoms would be frozen (activated) as two Rydberg atoms are in the same (different) ground states. Based on URP, one can achieve a three-qubit controlled-phase gate, steady-state entanglement, and autonomous quantum error correction.

In this paper, we modify the original URP mechanism by acting on different atoms with different frequencies of the driving field, respectively. Such a modified URP mechanism can directly implement a three-qubit controlled-not (Toffoli) gate without using the synthesis method of Hadamard gate plus controlled-phase gate~\cite{Isenhower2010PRL,Fedorov2012Nature,Eastin2013PRA,Zahedinejad2015PRL,Maslov2016PRA,KYH2018PRA,Levine2019PRL,WJL2019JPA,WJL2019EPL,KYH2020PRA_1,SZC2020PRA,KYH2020PRA_2,Rasmussen2020PRA}. Meanwhile, our scheme is not only immune to the variation of RRI strength between the control atoms, but also has a certain robustness to the fluctuation of the interaction strength between the control atoms and the target atom. And our scheme can be further extended to the four-qubit case without altering the operating time.

\section{Implementation of the Toffoli gate}\label{II}
\begin{figure}[ht]
	\centering
	\includegraphics[width=\linewidth]{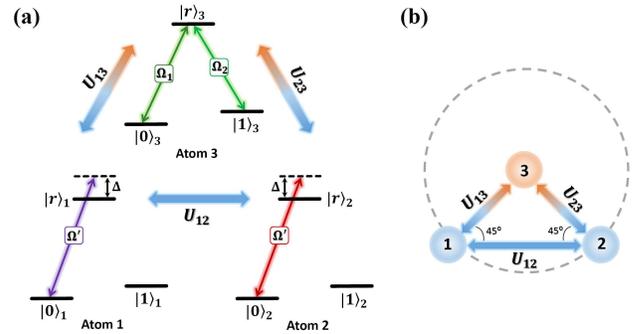}\\
	\caption{(a) The diagram of atom level configuration. $|0\rangle_{m}$ and $|1\rangle_{m}$ are ground states, and $|r\rangle_{m}$ is excited Rydberg state, $m=\left\lbrace1,~2,~3\right\rbrace$. Atom~1 and Atom~2 are control qubits, driven by a classical field of Rabi frequency $\Omega'$ with blue detuning $\Delta$, respectively. Atom~3 is target qubit driven by classical fields with Rabi frequencies $\Omega_1$ and $\Omega_2$. $U_{12}$, $U_{13}$ and $U_{23}$ denote the RRI strengths. (b) Schematic representation of three interacting Rydberg atoms.}\label{Fig1}
\end{figure}

\noindent The Toffoli gate can carry out the flipping of target qubit between $|0\rangle$ and $|1\rangle$ if and only if control qubits are both in $|1\rangle$. It can be expressed as follows
\begin{align}\label{e1}
U_{\textrm{Toffoli}}=\sum^{1}_{i,j,k=0}|i,j,ij\oplus k\rangle \langle i,j,k|,
\end{align}
where $ij\oplus k=(i\times j+k)$ mod 2. In order to realize this operation, we consider a system composed of three Rydberg atoms, and the relevant configuration of the atomic level is illustrated in Fig.~1(a). Atom~1 and Atom~2 are control qubits which consist of ground states $|0\rangle$ and $|1\rangle$ and an excited Rydberg state $|r\rangle$. There is only one off-resonant transition between $|0\rangle$ and $|r\rangle$ with blue detuning $\Delta$ driven by a classical field of Rabi frequency $\Omega'$. We label the target qubit as Atom~3, and there are two resonant transitions between $|0\rangle\leftrightarrow|r\rangle$ and $|1\rangle \leftrightarrow|r\rangle$ with Rabi frequencies $\Omega_1$ and $\Omega_2$, respectively. The atomic alignment is shown in Fig.~1(b), the control qubits are on a dotted circle centered on the target qubit. In the interaction picture, the Hamiltonian of the total system can be written as~($\hbar=1$)
\begin{align}\label{e2}
H_{\textit{I}}&=\Omega'e^{i\Delta t}(|0\rangle_1\langle r|+|0\rangle_2\langle r|)
+\Omega_1|0\rangle_3\langle r|+\Omega_2|1\rangle_3\langle r|\nonumber\\
&~~~~+\mathrm{H.c.}+\sum_{j\neq k}U_{jk}|rr\rangle_{jk}\langle rr|,
\end{align}
where $U_{jk}$ is considered as van der Waals interaction strength $U_{\textrm{vdW}}=C_6/R^6$ between the $j$th and $k$th atom. $R$ is the distance between two Rydberg atoms and $C_6$ depends on the quantum numbers of the Rydberg state. To embody the effect of the RRI, we extend the Hamiltonian to a three-atom basis form and rewrite $H_{\textit{I}}$ after moving to a rotating frame with respect to $U_0=\exp(-{it}\sum_{j\neq k} U_{jk}|rr\rangle_{jk}\langle rr|)$ by using the formula $i\dot{U}_{0}^{\dagger}U_{0}+U_{0}^{\dagger}H_{I}U_{0}$. Then we have
	
\begin{align}\label{e3}
H_{\textit{IR}}&=\sum_{\alpha,\beta=0}^{1}\Omega_1(|\alpha\beta 0\rangle\langle \alpha\beta r|
+e^{-i\Delta t}|\alpha r0\rangle\langle \alpha rr|\nonumber\\
&~~~~+e^{-i\Delta t}|r\alpha0\rangle\langle r\alpha r|
+e^{-2i\Delta t}|rr0\rangle\langle rrr|)\nonumber\\
&~~~~+\Omega_2(|\alpha\beta1\rangle\langle \alpha\beta r|+e^{-i\Delta t}|\alpha r1\rangle\langle \alpha rr|\nonumber\\
&~~~~+e^{-i\Delta t}|r\alpha1\rangle\langle r\alpha r|+e^{-2i\Delta t}|rr1\rangle\langle rrr|)\nonumber\\
&~~~~+\Omega'(e^{i\Delta t}|0\alpha\beta\rangle\langle r\alpha\beta|+|0\alpha r\rangle\langle r\alpha r|\nonumber\\
&~~~~+e^{i(\Delta-U_{12})t}|0r\alpha\rangle\langle rr\alpha|
+e^{-iU_{12}t}|0rr\rangle\langle rrr|\nonumber\\
&~~~~+e^{i\Delta t}|\alpha0\beta\rangle\langle \alpha r\beta|
+|\alpha 0r\rangle\langle \alpha rr|\nonumber\\
&~~~~+e^{i(\Delta-U_{12})t}|r0\alpha\rangle\langle rr\alpha|
+e^{-iU_{12}t}|r0r\rangle\langle rrr|)\nonumber\\
&~~~~+\mathrm{H.c.},
\end{align}
where we have utilized the URP condition as $U_{13}=U_{23}=\Delta$. Under the condition $\Delta\gg\left\lbrace\Omega', \Omega_1, \Omega_2\right\rbrace$, we can further simplify Eq.~(3) by eliminating the high-frequency oscillating terms and obtain

\begin{align}\label{e4}
H'_{\textit{IR}}=H^{(1)}_{\textit{IR}}+H^{(2)}_{\textit{IR}}+H^{(3)}_{\textit{IR}}+H^{(4)}_{\textit{IR}},
\end{align}
where
\begin{align}
H^{(1)}_{\textit{IR}}&=\Omega_1|000\rangle\langle 00r|+\Omega_2|001\rangle\langle 00r|
+\Omega'(|00r\rangle\langle r0r|\nonumber\\
&~~~~+|00r\rangle\langle 0rr|+e^{-iU_{12}t}|0rr\rangle\langle rrr|\nonumber\\
&~~~~+e^{-iU_{12}t}|r0r\rangle\langle rrr|)+\mathrm{H.c.},\nonumber\\
H^{(2)}_{\textit{IR}}&=\Omega_1|010\rangle\langle 01r|+\Omega_2|011\rangle\langle 01r|
+\Omega'|01r\rangle\langle r1r|\nonumber\\
&~~~~+\mathrm{H.c.},\nonumber\\
H^{(3)}_{\textit{IR}}&=\Omega_1|100\rangle\langle 10r|+\Omega_2|101\rangle\langle 10r|
+\Omega'|10r\rangle\langle 1rr|\nonumber\\
&~~~~+\mathrm{H.c.},\nonumber\\
H^{(4)}_{\textit{IR}}&=\Omega_1|110\rangle\langle 11r|+\Omega_2|111\rangle\langle 11r|+\mathrm{H.c.}.\nonumber
\end{align}
The transitions of different subspaces are described from $H^{(1)}_{\textit{IR}}$ to $H^{(4)}_{\textit{IR}}$ in Eq.~(4), respectively. It includes both the desired and undesired transitions for implementation of the Toffoli gate. In Fig. 2(a), we plot the transition process of computational basis states $\left\lbrace |000\rangle, |001\rangle \right\rbrace$ dominated by $H^{(1)}_{\textit{IR}}$. When we select the condition  $U_{12}\gg\Omega'$, the transitions from $|r0r\rangle(|0rr\rangle)$ to $|rrr\rangle$ can be eliminated as the high-frequency oscillating terms in $H^{(1)}_{\textit{IR}}$. Then the subspace defined by $H^{(1)}_{\textit{IR}}$ is reduced from six dimensions to five dimensions, which is similar to the process described by $H^{(2)}_{\textit{IR}}$ and $H^{(3)}_{\textit{IR}}$. If we further assume the condition $\Omega'\gg\left\lbrace \Omega_1, \Omega_2 \right\rbrace $, the strong interactions in $H^{(1)}_{\textit{IR}}$ can be diagonalized as $\sum_{p=\pm, 0}\lambda_p|\Phi_{p}\rangle\langle \Phi_{p}|$ with $\lambda_{\pm}=\pm\sqrt{2}\Omega'$ and $\lambda_{0}=0$, where the dressed states $|\Phi_{\pm}\rangle=(\sqrt{2}|00r\rangle\pm|r0r\rangle\pm|0rr\rangle)/2$ and $|\Phi_{0}\rangle=(|r0r\rangle-|0rr\rangle)/\sqrt{2}$. Thus, the effective transitions between the computational basis states $\{|000\rangle, |001\rangle\}$ and $|\Phi_{\pm}\rangle$ can be represented as $(\Omega_1|000\rangle+\Omega_2|001\rangle)(e^{-\sqrt{2}i\Omega' t}\langle \Phi_{+}|+e^{\sqrt{2}i\Omega' t} \langle \Phi_{-}|)/\sqrt{2}+\mathrm{H.c.}$ as shown in Fig.~2(b), from which we can see that the interactions between $\left\lbrace|000\rangle, |001\rangle \right\rbrace$ and $\left\lbrace |\Phi_{+}\rangle, |\Phi_{-}\rangle \right\rbrace$ are largely detuned under the condition $|\lambda_{\pm}|\gg\left\lbrace\Omega_1, \Omega_2 \right\rbrace$ owing to $\Omega'\gg\left\lbrace\Omega_1, \Omega_2 \right\rbrace$. At the same time, there is neither a transition between the ground states, nor Stark shifts of the ground states, because these transition paths mediated by the independent channels $|\Phi_{\pm}\rangle$ interfere destructively. In addition, the evolutions of other computational basis states $\left\lbrace |010\rangle, |011\rangle \right\rbrace$ in $H^{(2)}_{\textit{IR}}$ and $\left\lbrace |100\rangle, |101\rangle \right\rbrace$ in $H^{(3)}_{\textit{IR}}$ are similar to the above case as $U_{12}\gg\Omega'$, thence these transition processes can also be forbidden.

\begin{figure}[ht]
	\centering
	\includegraphics[width=\linewidth]{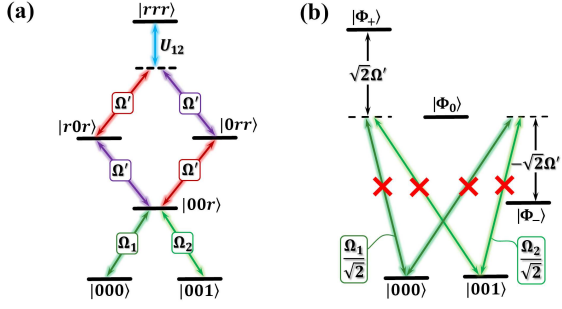}\\
	\caption{(a) The level transitions between ground states $|000\rangle$ and $|001\rangle$ and excited Rydberg states under URP condition as $U_{13}=U_{23}=\Delta\gg\Omega'\gg\left\lbrace\Omega_1, \Omega_2 \right\rbrace$. (b) The effective transitions between ground states and dressed states under the condition $U_{12}\gg\Omega'$ and other situations are the same with above. The red cross marks indicate that the transition processes are forbidden.}\label{Fig2}
\end{figure}

For a more general case of $U_{12}$ which is not much larger than $\Omega'$, we can use the effective operator method~\cite{Kastoryano2011PRL,Reiter2012PRA,SLT2011PRA} to obtain a result similar to the above case, i.e., $|000\rangle$ and $|001\rangle$ will not evolve. The time-independent form of $H^{(1)}_{\textit{IR}}$ in Eq.~(4) can be expressed as $H^{(1)}_{\textit{IR}}=H_{\textit{e}}+V_{+}+V_{-}$, where $H_{\textit{e}}=\Omega'(|00r\rangle\langle r0r|+|00r\rangle\langle 0rr|+|0rr\rangle\langle rrr|+|r0r\rangle\langle rrr|)+\mathrm{H.c.}+U_{12}|rrr\rangle\langle rrr|$ and $V_{-}=V_{+}^{\dagger}=\Omega_1|000\rangle \langle 00r|+\Omega_2|001\rangle \langle 00r|$.  Under the condition that the Rabi frequencies $\Omega_1$ and $\Omega_2$ are sufficiently weak compared with $\Omega'$ and $U_{12}$, the effective Hamiltonian of subspace $\left\lbrace |000\rangle, |001\rangle \right\rbrace$ can be described as
\begin{align}\label{e5}
H^{{\textit{IR}}{(1)}}_{\mathrm{eff}}&=-[V_{-} H_{\textit{e}}^{-1} V_{+} + V_{-} (H_{\textit{e}}^{-1})^{\dagger} V_{+}]/2\nonumber\\
&=-[\Omega_1^2|000\rangle \langle 000|+\Omega_2^2|001\rangle \langle 001|\nonumber\\
&~~~~+\Omega_1\Omega_2(|000\rangle \langle 001|+|001\rangle \langle 000|)]/U_{12},
\end{align}
where the inverse matrix $H^{-1}_{\textit{e}}$ can be calculated by replacing 0 on the diagonal element of the matrix $H_{\textit{e}}$ with an infinitesimal in order to avoid singular value. This equation does not involve the transitions from the ground states to the Rydberg state, but two additional Stark-shift terms, which is weak enough compared to $\left\lbrace \Omega_1, \Omega_2\right\rbrace$ in Eq.~(4) to be negligible. As a result, considering the above two magnitude of $U_{12}$, the effective Hamiltonian of the total system can be simplified as
\begin{align}\label{e6}
H_{\textrm{eff}}\simeq H^{(4)}_{\textit{IR}}=\Omega_1|110\rangle\langle 11r|+\Omega_2|111\rangle\langle 11r|+\mathrm{H.c.}.
\end{align}
Thus the Toffoli gate can be carried out within one step as $t=\pi/\Omega$ by setting $\Omega=\sqrt{\Omega_1^2+\Omega_2^2}$ and $\Omega_1=-\Omega_2$. It is worth noting that there is no two-excited Rydberg state in the effective Hamiltonian, so our scheme does not need to strictly control the interaction strength between Rydberg atoms like Rydberg antiblockade. As a result, the energy level shift caused by dipole-dipole force arising from the interaction between two Rydberg states is avoided effectively, and the corresponding Rydberg state will not be influenced by this factor.

\section{Numerical simulation}\label{III}
\begin{figure}[ht]
	\centering
	\includegraphics[width=\linewidth]{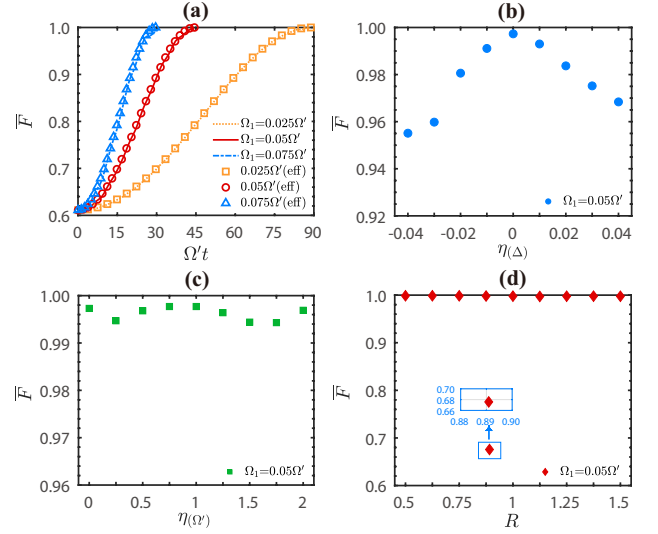}\\
	\caption{(a) The average fidelities of the Toffoli gate governed by full (effective) Hamiltonian with different Rabi frequencies $\Omega_1=\left\lbrace 0.025\Omega', 0.05\Omega', 0.075\Omega'\right\rbrace$, where $U_{13}=U_{23}=\Delta$, $U_{12}=\Delta/8$, and $\Delta=50\Omega'$. (b) The variation tendency of the average fidelities with different mismatching rate $\eta_{\Delta}$ between $U$ and $\Delta$, where $U_{13}=U_{23}=U$, $\eta_{\Delta}=(U-\Delta)/\Delta$, $U_{12}=\Delta/8$, $\Delta=50\Omega'$, and $\Omega_1=0.05\Omega'$. (c) The variation trend of the average fidelities with different mismatching rate $\eta_{(\Omega')}=(\Omega''-\Omega')/\Omega'$, where $\Omega''$ is the Rabi frequency driven the control qubit Atom~2 and the other parameters are the same as (a) when $\Omega_1=0.05\Omega'$. (d) The variation trend of the average fidelities with different distances between control qubits when $\Omega_1=0.05\Omega'$, $U_{13}=U_{23}=\Delta$, and $\Delta=50\Omega'$, where the value of distance $R$ corresponds to $U_{12}=\Delta$. The insets show the structures of different distances and the enlarged view of the average fidelity for $U_{12}=2\Delta$, respectively.}\label{Fig3}
\end{figure}

\noindent To assess the performance of the Toffoli gate, we introduce the trace-preserving quantum-operator-based (TPQO) average fidelity method defined as~\cite{Nielsen2002PLA}
\begin{align}\label{e7}
\overline{F}(\varepsilon,\mathcal{O})=\frac{\sum_{j}\operatorname{tr}\left(\mathcal{O} \mathcal{O}_{j}^{\dagger} {\mathcal{O}}^{\dagger} \varepsilon\left(\mathcal{O}_{j}\right)\right)+d^{2}}{d^{2}(d+1)},
\end{align}
where $\mathcal{O}_j$ is the tensor of Pauli matrices $III$, $IIX$, ... , $ZZZ$, $\mathcal{O}$ is the perfect Toffoli gate, $d=8$ for a three-qubit quantum logic gate, and $\varepsilon$ is the trace-preserving quantum operation achieved through our scheme. By using the TPQO average fidelity method, we plot average fidelities of the Toffoli gate with different Rabi frequencies $\Omega_1$ in Fig.~3(a), where $U_{13}=U_{23}=\Delta$, $U_{12}=\Delta/8$ and $\Delta=50\Omega'$ by using the full Hamiltonian $H_{\textit{I}}$ in Eq.~(2) and the effective Hamiltonian $H_{\textrm{eff}}$ in Eq.~(6), respectively. And the results of $H_{\textit{I}}$ are in good agreement with $H_{\textrm{eff}}$. The average fidelity can reach at 0.998, 0.9972 and 0.9914 with the increase of Rabi frequency $\Omega_1$ from $0.025\Omega'$ to $0.075\Omega'$. From this we can see that the larger value of $\Omega_1$, the worse the approximation. Therefore, without considering any decohenrece, we will get a higher fidelity if the value of $\Omega_1$ is relatively small.

Next, we take into account the situation that $U_{13}(U_{23})$ and $\Delta$ are not strictly equal. Compared with Rydberg antiblockade that requires exact control of the interaction strength between Rydberg atoms, our scheme does not need to strictly control the interaction strength between the control atoms and the target atom. In Fig.~3(b), we depict the average fidelities with different mismatching rate $\eta_{(\Delta)}$ between $U$ and $\Delta$, where $U_{13}=U_{23}=U$ and $\eta_{(\Delta)}=(U-\Delta)/\Delta$, $U_{12}=\Delta/8$, $\Delta=50\Omega'$ and $\Omega_1=0.05\Omega'$. The results show that the fidelities can still be maintained above 0.98 when $\eta_{(\Delta)}$ is from $-0.02$ to $0.02$, which is equivalent to the difference between $U$ and $\Delta$ can reach the order of magnitude $\Omega'$. Thus, even if the matching relationship between $U$ and $\Delta$ is slightly off, the fidelity is still high and has little impact on the implementation of the gate. Moreover, we discuss the influence of the gate fidelity based on the strengths of Rabi frequencies driven the control qubits. In Fig.~3(c), we suppose that the control qubit Atom~2 is driven by Rabi frequency $\Omega''$ and plot the average fidelities with different mismatching rate $\eta_{(\Omega')}=(\Omega''-\Omega')/\Omega'$ between $\Omega'$ and $\Omega''$ when $U_{13}=U_{23}=\Delta$, $U_{12}=\Delta/8$, $\Delta=50\Omega'$ and $\Omega_1=0.05\Omega'$. It can be seen that the fidelities with different $\Omega''$ are all remained above 0.994. Hence, the Rabi frequencies driven control qubits does not have to be exactly equal when they are much greater than $\left\lbrace \Omega_1, \Omega_2\right\rbrace$.

In addition, we would like to study the impact of changes in the value of $U_{12}$ on the scheme. In Fig~3(d), we plot the average fidelities with different distance between control qubits when $\Omega_1=0.05\Omega'$ under the conditions of $U_{13}=U_{23}=\Delta$ and $\Delta=50\Omega'$. The distance between control atoms $1R$ corresponds to the relation $U_{12}=\Delta=50\Omega'$. As the distance increases from $0.5R$ to $1.5R$, $U_{12}$ can be changed from $3200\Omega'$ to $4.39\Omega'$, but the corresponding average fidelities are remained above 0.997. Thus, the magnitude of $U_{12}$ hardly influences the implementation of the Toffoli gate. However, we notice a special case where the distance is approximately equal to $0.8909R$, i.e. $U_{12}=2\Delta$. At this point, the average fidelity can only reach 0.6757. The reason is that the two-photon resonance under $U_{12}=2\Delta$ results in the direct transition between $|000\rangle(|001\rangle)$ and $|rr0\rangle(|rr1\rangle)$, which affects the final average fidelity. And the transitions of $|0r0\rangle(|r00\rangle)\leftrightarrow|rr0\rangle$ and $|0r1\rangle(|r01\rangle)\leftrightarrow|rr1\rangle$ in Eq.~(3) are decoupled from ground states and can be omitted in most cases except $U_{12}=2\Delta$. Thus, the value of $U_{12}$ should keep away from the vicinity of $2\Delta$ when $U_{12}\gg\Omega'$.

\begin{figure}[ht]
	\centering
	\includegraphics[width=\linewidth]{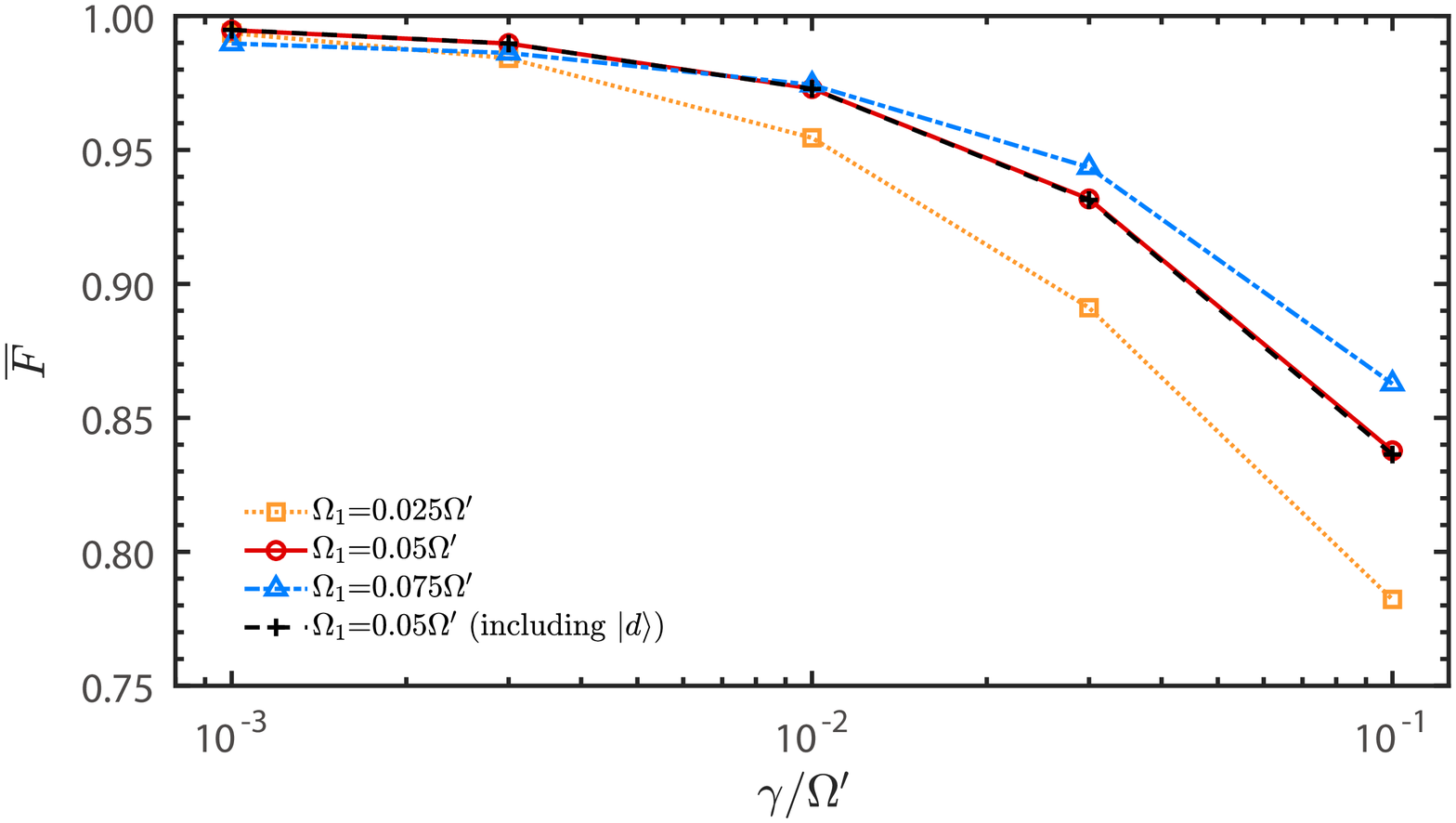}\\
	\caption{The variation tendency of the fidelities of Toffoli gate scheme with different Rabi frequencies $\Omega_1=\left\lbrace 0.025\Omega', 0.05\Omega', 0.075\Omega'\right\rbrace$ versus decay rate $\gamma$ and the other parameters are same as Fig.~3(a).}\label{Fig4}
\end{figure}

In the presence of the spontaneous emission of Rydberg atoms, the system can be dominated by the following master equation
\begin{align}\label{e8}
\dot{\rho}=-i[H_{\textit{I}},\rho]+\frac{\gamma}{2}\sum^6_{j=1}\left(2\sigma_{j}\rho\sigma_{j}^{\dagger}
-\sigma_{j}^{\dagger}\sigma_{j}\rho
-\rho\sigma_{j}^{\dagger}\sigma_{j}\right),
\end{align}
where $\gamma$ is the decay rate of the Rydberg state, and we have assumed that the branching ratios of spontaneous emission from Rydberg state $|r\rangle$ to the ground states $|0\rangle$ and $|1\rangle$ are both $\gamma/2$, $\sigma_{1}=|0\rangle_{1}\langle r|$, $\sigma_{2}=|1\rangle_{1}\langle r|$, $\sigma_{3}=|0\rangle_{2}\langle r|$, $\sigma_{4}=|1\rangle_{2}\langle r|$, $\sigma_{5}=|0\rangle_{3}\langle r|$ and $\sigma_{6}=|1\rangle_{3}\langle r|$ denote six decay channels of the three atoms, respectively. In Fig.~4, we plot the variation trend of the average fidelities under the influence of decay rate $\gamma$ with different $\Omega_1$ from $0.025\Omega'$ to $0.075\Omega'$ under the conditions of $U_{13}=U_{23}=\Delta$, $U_{12}=\Delta/8$ and $\Delta=50\Omega'$. It can be seen that although the fidelity of $\Omega_1=0.075\Omega'$ is lower than others when $\gamma=0.001\Omega'$, and the decreasing of the fidelity is less than others with the increasing of the decay rate. Thus, with the appropriate increase of Rabi frequency, the interaction time is shortened, thereby further reducing the influence of dissipation on the scheme. In addition, there is a probability that atoms in the excited Rydberg states will spontaneously decay into an external level out of $\{|0\rangle, |1\rangle\}$. To study this situation, we incorporate an uncoupled state $|d\rangle$ in {Eq.~(8)}, and suppose the branching ratio of spontaneous emission from $|r\rangle$ to $|d\rangle$ is equal to the branching ratio to the computational basis states $|0\rangle$ and $|1\rangle$. In Fig.~4, we find the variation trend of the fidelities under the influence of $\gamma$ is consistent with the case without considering state $|d\rangle$ when $\Omega_1=0.05\Omega'$. Therefore, it has little influence on our scheme whether the state outside computational basis states exists or not.

\begin{table}[ht]
	\centering
	\setlength{\abovecaptionskip}{2pt}
	\setlength{\belowcaptionskip}{2pt}
	\caption{The average fidelities of the Toffoli gate under different experimental parameters. The variable parameters are $U_{12}$=$2\pi\times (50/8, 50, 50/64)$ MHz, $U$=$2\pi\times (49.5, 50, 50.5)$ MHz and $\Omega''$=$2\pi\times (1,2)$ MHz. Other same parameters are selected as $(\Omega_1,~\Omega',~\Delta)=2\pi\times (0.05,~1,~50)$ MHz and $\gamma$= 3.125~kHz.}
	\begin{tabular}{cccc}
		\hline\hline
		~~~$U_{12}/2\pi$~(MHz)~~~&~~~$U/2\pi$~(MHz)~~~&~~~$\Omega''/2\pi$~(MHz)~~~&~~~$\overline{F}$~~~\\
		\hline
		50/8~~~&~~~50~~~~~~&~~~1~~~&~~~0.9961~~~\\
		50/8~~~&~~~50~~~~~~&~~~2~~~&~~~0.9951~~~\\
		50/8~~~&~~~49.5~~~&~~~1~~~&~~~0.99~~~~~~~\\
		50/8~~~&~~~49.5~~~&~~~2~~~&~~~0.9918~~~\\
		50/8~~~&~~~50.5~~~&~~~1~~~&~~~0.9931~~~\\
		50/8~~~&~~~50.5~~~&~~~2~~~&~~~0.9934~~~\\
		50~~~&~~~50~~~~~~&~~~1~~~&~~~0.9965~~~\\
		50~~~&~~~49.5~~~&~~~2~~~&~~~0.9919~~~\\
		50~~~&~~~50.5~~~&~~~1~~~&~~~0.9936~~~\\	
		50/64~~~&~~~50~~~~~~&~~~2~~~&~~~0.9949~~~\\
		\hline\hline
	\end{tabular}
\end{table}

In experiment, we choose Rydberg atom with the appropriate principal quantum number to maintain a relatively long radiative lifetime $\tau$ for suppressing the influence of spontaneous emission. Therefore, we use the Rydberg state $|97d_{5/2}, m_j=5/2\rangle $ ($|C_6|$ = 37.6946~Thz$\cdot\mu m^6$) of $^{87}\textrm{Rb}$ atom with the lifetime $\sim 320~\mu s$ and decay rate $\gamma\simeq$ 3.125~kHz ($\gamma\simeq1/\tau$) referring to prior works~\cite{SSL2017PRA_1,Isenhower2010PRL,ZXL2010PRA,ZXL2012PRA,QJ2012PRA,Beguin2013PRL,Weber2017JPB}. And the required strength of Rabi frequency excited to the Rydberg state can be continuously adjusted to $2\pi\times10$ MHz~\cite{SSL2018PRA,deLeseleuc2018PRA}. According to the atomic alignment in Fig.~1(b) ($U_{12}=2\pi\times 50/8$ MHz), we consider the influence of different $U$ and $\Omega''$ on the scheme, and the corresponding values are listed in Table 1, $U$=$2\pi\times (49.5, 50, 50.5)$ MHz and $\Omega''$=$2\pi\times (1,2)$ MHz. If the arrangement of the three atoms is an equilateral triangle ($U_{12}=2\pi\times 50$ MHz) or a straight line ($U_{12}=2\pi\times 50/64$ MHz), we can also obtain the corresponding fidelities listed in Table 1. Other same parameters are selected as $(\Omega_1,~\Omega',~\Delta)=2\pi\times (0.05,~1,~50)$ MHz and $\gamma$= 3.125~kHz.

\section{Four-qubit case}\label{IV}
\begin{figure}[ht]
	\centering
	\includegraphics[width=\linewidth]{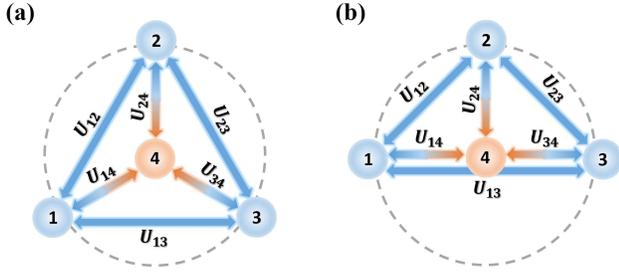}\\
	\caption{(a) Schematic representation of four interacting Rydberg atoms under the same distance among control qubits. (b) Schematic representation under the different distance among control qubits. The distances between target qubit and control qubits are all the same.}\label{Fig5}
\end{figure}

\begin{figure}[ht]
	\centering
	\includegraphics[width=\linewidth]{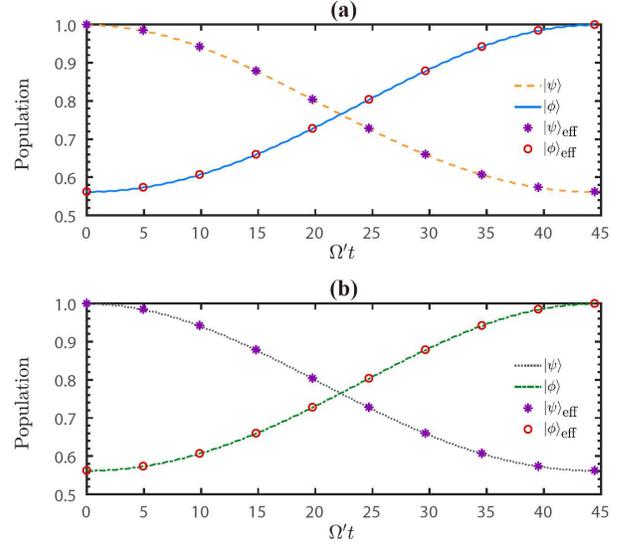}\\
	\caption{(a) The population of state $|\phi\rangle$ and $|\psi\rangle$ governed by full and effective Hamiltonian of four-qubit Toffoli gate with $U_{12}=U_{13}=U_{23}=\Delta/27$. (b) The population of state $|\phi\rangle$ and $|\psi\rangle$ governed by full and effective Hamiltonian of four-qubit Toffoli gate with $U_{13}=\Delta/64$ and $U_{12}=U_{23}=\Delta/8$. The other same parameter conditions are $U_{14}=U_{24}=U_{34}=\Delta$, $\Delta=50\Omega'$ and $\Omega_1=0.05\Omega'$.}\label{Fig6}
\end{figure}

\noindent Based on the above analysis, our scheme can be directly extended to implement the four-qubit Toffoli gate. Here we consider two structures as shown in Fig.~5(a) and~5(b) where Atom~1,~2,~3 are control qubits. The target qubit is labeled as Atom~4. The Hamiltonian of the four-atom system can be represented as
\begin{align}\label{e9}
H'_{\textit{I}}&=\Omega'e^{i\Delta t}(|0\rangle_1\langle r|+|0\rangle_2\langle r|+|0\rangle_3\langle r|)
+\Omega_1|0\rangle_4\langle r|\nonumber\\
&~~~~+\Omega_2|1\rangle_4\langle r|+\mathrm{H.c.}+\sum_{m\neq n}U_{mn}|rr\rangle_{mn}\langle rr|,
\end{align}
where we have set the URP conditions for the four-qubit system as $U_{14}=U_{24}=U_{34}=\Delta\gg\Omega'\gg\left\lbrace\Omega_1, \Omega_2\right\rbrace$ and $\left\lbrace U_{12}, U_{13}, U_{23}\right\rbrace\neq 2\Delta$. Then, we can obtain the effective Hamiltonian of the four-qubit system
\begin{align}\label{e10}
H'_{\textrm{eff}}=\Omega_1|1110\rangle\langle 111r|+\Omega_2|1111\rangle\langle 111r|+\mathrm{H.c.}.
\end{align}
In order to assess the performance of the four-qubit Toffoli gate, we select $|\psi\rangle=(\sum\nolimits^1_{i,j,k,l=0}|ijkl\rangle-2|1110\rangle)/4$ as the initial state, $|\phi\rangle=(\sum\nolimits^1_{i,j,k,l=0}|ijkl\rangle-2|1111\rangle)/4$ as the ideal finial state and observe the transfer of population between states $|\psi\rangle$ and $|\phi\rangle$. According to the atomic alignments in Fig.~5(a) and~5(b), we respectively plot the population of $|\phi\rangle$  with $H'_{\textrm{I}}$ in Eq.~(9) under condition $U_{14}=U_{24}=U_{34}=\Delta$, $U_{12}=U_{23}=U_{13}=\Delta/27$ in Fig.~6(a), and the population of $|\psi\rangle$ with $H'_{\textrm{eff}}$ in Eq.~(10) under condition $U_{14}=U_{24}=U_{34}=\Delta$, $U_{13}=\Delta/64$, $U_{12}=U_{23}=\Delta/8$ in Fig.~6(b). These results show that the populations are almost completely transferred, and the tendencies of the populations governed by the full Hamiltonian and the effective Hamiltonian are identical. Therefore, as long as the corresponding URP conditions are satisfied, the four-qubit Toffoli gate can be implemented in one step without altering the operating time. However, it is currently difficult to extend the scheme to more than four qubits since the interactions become complicated with increase of the number of qubits. We look forward to coming up with a simpler scheme to implement multiqubit Toffoli gates in the follow-up work.

\section{Summary}\label{V}
In summary, our work has provided a scheme to implement a Toffoli gate within one step based on unconventional Rydberg pumping (URP) mechanism. Compared with Rydberg antiblockade, our scheme does not need to strictly control the interaction strength between atoms. The fluctuations of parameters $U_{12}$, $U$ and $\Omega''$ are allowed within a certain range for maintaining a high fidelity of gate. Meanwhile, the scheme can be directly extended to the four-qubit case without changing the operation time. We hope that our scheme can provide a new approach for the implementation of scalable Rydberg quantum gates.

\section{Acknowledgements}
This work is supported by National Natural Science Foundation of China (11774047); Fundamental Research Funds for the Central Universities (2412020FZ026); Natural Science Foundation of Jilin Province (JJKH20190279KJ).

\bibliography{pra}

\end{document}